%% file: main.tex
\documentclass[10pt,conference]{IEEEtran}

\usepackage[utf8]{inputenc}
\usepackage{xcolor}
\usepackage{multirow}
\usepackage{tabularx}
\usepackage{makecell}
\usepackage{graphicx}
\usepackage{amsmath}
\usepackage[flushleft]{threeparttable}
\usepackage{subcaption}
\usepackage{enumitem}
\usepackage{hyperref}

\definecolor{pink}{rgb}{0.9,0,0.9}

\newcolumntype{L}{>{\raggedright\arraybackslash}X}
\newcolumntype{P}[1]{>{\centering\arraybackslash}p{#1}}

\begin{document}

\title{Analysis and Detection of Information Types of Open Source Software Issue Discussions}


\author{
\IEEEauthorblockN{Deeksha Arya\IEEEauthorrefmark{1}, Wenting Wang\IEEEauthorrefmark{1}, Jin L.C. Guo\IEEEauthorrefmark{1}, Jinghui Cheng\IEEEauthorrefmark{2}}
\IEEEauthorblockA{\IEEEauthorrefmark{1}School of Computer Science, McGill University, Montreal, Canada}
\IEEEauthorblockA{\IEEEauthorrefmark{2}Department of Computer and Software Engineering, Polytechnique Montreal, Montreal, Canada}
\IEEEauthorblockA{\IEEEauthorrefmark{1}\emph{\{deeksha.arya, wenting.wang\}@mail.mcgill.ca}, \emph{jguo@cs.mcgill.ca} \IEEEauthorrefmark{2}\emph{jinghui.cheng@polymtl.ca}}
}

\maketitle

\begin{abstract}
Most modern Issue Tracking Systems (ITSs) for open source software (OSS) projects allow users to add comments to issues. Over time, these comments accumulate into discussion threads embedded with rich information about the software project, which can potentially satisfy the diverse needs of OSS stakeholders. However, discovering and retrieving relevant information from the discussion threads is a challenging task, especially when the discussions are lengthy and the number of issues in ITSs are vast. In this paper, we address this challenge by identifying the information types presented in OSS issue discussions. Through qualitative content analysis of 15 complex issue threads across three projects hosted on GitHub, we uncovered 16 information types and created a labeled corpus containing 4656 sentences. Our investigation of supervised, automated classification techniques indicated that, when prior knowledge about the issue is available, Random Forest can effectively detect most sentence types using conversational features such as the sentence length and its position. When classifying sentences from new issues, Logistic Regression can yield satisfactory performance using textual features for certain information types, while falling short on others. Our work represents a nontrivial first step towards tools and techniques for identifying and obtaining the rich information recorded in the ITSs to support various software engineering activities and to satisfy the diverse needs of OSS stakeholders.

\begin{IEEEkeywords}
	collaborative software engineering, issue tracking system, issue discussion analysis
\end{IEEEkeywords}

\end{abstract}

\input{introduction.tex}
\input{related_work.tex}

\input{issue_comment_coding.tex}
\input{use_case.tex}
\input{supervised_classification.tex}
\input{evaluation.tex}
\input{threats_to_validity.tex}
\input{conclusion.tex}

\section{Acknowledgments}
This work was partially funded by the Canada NSERC Dicovery Grant RGPIN-2018-04470. We also thank Michalis Famelis for his early feedback to our work.

\bibliographystyle{abbrv}

\end{document}

%% file: introduction.tex
\section{Introduction}
\label{sec:intro}
Software development teams often use Issue Tracking Systems (ITSs) to manage affairs or cases during the development process. The issues managed by ITSs include bug reports, new feature requests, enhancements, documentation updates, general tasks to be completed, or even feedback solicitations on rough ideas. Such issue management platforms play an essential role in supporting software engineering activities such as bug triaging, impact analysis, and release planning. Additionally, they act as a rich source of information about the software project for developers, users, and other stakeholders. For Open Source Software (OSS) projects in particular, ITSs (e.g. the \textit{Issues} feature of \href{https://github.com}{GitHub}) are critical in engaging various stakeholders throughout the project life-cycle.

Once an issue is submitted, OSS stakeholders are normally able to follow up by leaving comments. These comments serve as a mechanism for asynchronous conversation among different stakeholders around the target issues. For example, comments may be associated with issue severity, tentative design, implementation details, and references to similar work, among other discussions. Users may also express their willingness to contribute to the project, report progress, or ask for feedback and help with respect to matters at hand. Over time, these comments constitute discussion threads embedded with rich and abundant information about the associated issue and the project as a whole.

\begin{figure}[t]
\centering
\includegraphics[width=\columnwidth]{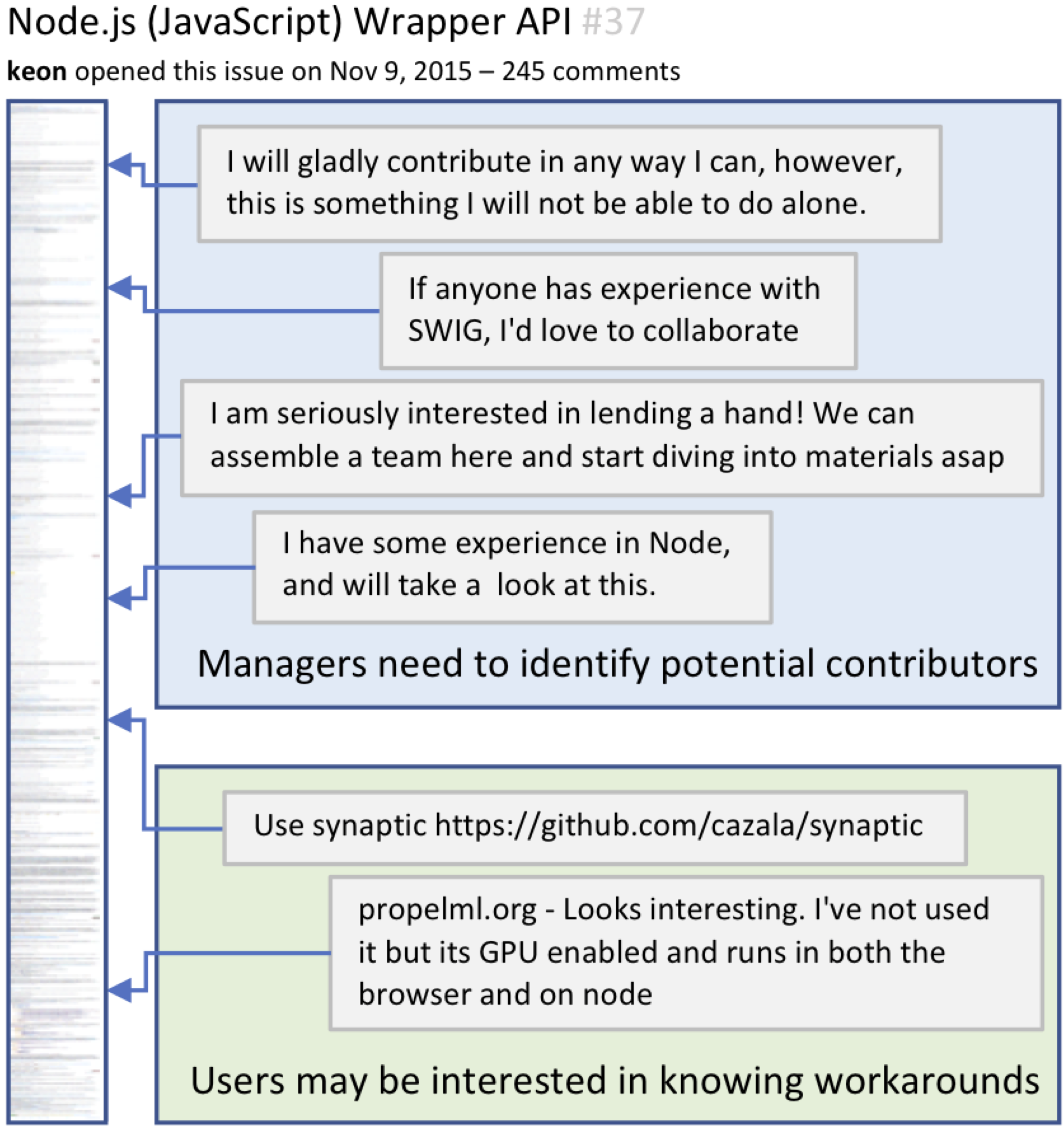}
\caption{Example issue discussion thread \textit{TensorFlow \#37} (compressed to the left), where useful information such as potential contributors and temporary workarounds is hidden in the 245 comments.}
\label{fig:motivation}
\end{figure}


Based on their end goals and current tasks, OSS project stakeholders generally try to acquire diverse information from the issue discussion threads. For example, project managers may want to see if any participants have expressed interest in contributing to solving the issue; they may also need to monitor the solution progress reported in the comments. On the other hand, potential contributors may want to look into the nature of the issue and the complexity of the solution proposed to estimate the required effort before making any commitment towards working on the project. An end user may want to look for a quick workaround to remove a roadblock in their own project before a complicated issue is resolved \cite{Cheng:2018:OSC:3170427.3188467}.

However, the issue discussion threads can become very lengthy and difficult to follow, especially for complex issues which may have a major impact on the project \cite{PR_controbutor}. Additionally, OSS stakeholders usually have mixed backgrounds and often need to undergo extensive discussions before reaching a common ground; this aggravates the problem of information overload in issue discussion threads. Our preliminary analysis on 82 GitHub projects revealed that on average, there are more than 170 issues per project that contain over 20 comments. In these cases, 10 participants on average take part in the discussion. These issues were also the ones that took longer to resolve and required stakeholders to constantly revisit. 
We have also observed that OSS participants often ask questions for which answers already exist in earlier comments. This phenomenon indicated the difficulty of finding relevant information from documented issue discussions, particularly for people who lack familiarity of the project and a proper understanding of the context. Figure \ref{fig:motivation} illustrates the diverse needs that can be fulfilled by the issue discussion threads and the considerable effort involved in obtaining such information.




Issue metadata, such as tags and labels, are designed to partially address the problem of information overload by allowing OSS stakeholders to quickly grasp the most important information such as the origin and the status of the issue. In practice, however, the metadata are often outdated, incomplete or even misleading \cite{Wang2012}. Even when managed diligently, it usually only represents the interests of the project management team or the core developers and thus is inadequate in fulfilling the needs of the diverse OSS project participants. As a result, it is imperative to develop techniques that take full advantage of the rich information embedded in ITSs to help various OSS stakeholders quickly acquire their desired information. 



In this work, we take the first step toward such techniques by aiming to understand and automatically identify the information types presented in comments of OSS issue discussions. Knowledge about these information types, their primary objectives, and their structures are critical for designing tools to support various OSS stakeholders consume information embedded in ITSs. In order to identify the principle information types from the discussions, we first followed a content analysis process to empirically examine 15 complex discussion threads (containing 4656 sentences in total) from three popular OSS projects hosted on GitHub, covering bug reports, enhancement proposals, and feature requests. We then evaluated the effectiveness of using supervised methods to automatically identify those information types in the issue comments. Our work addresses the following research questions:

\begin{itemize}[leftmargin=0.15in, label={}]
    \item \textbf{RQ1}: What are the main information types presented in issue discussions that can potentially satisfy the needs of different stakeholders?
    \item \textbf{RQ2}: To what extent can automated classification methods identify the information types of issue comments?
    \begin{itemize}[label={}]
        \item \textbf{RQ2.1}: What are the performance and trade-offs of using different automated classification methods?
    \end{itemize}
\end{itemize}
This paper makes the following contributions: (1) an extensive codebook detailing the guidelines for annotating information types in OSS issue discussion threads; (2) a corpus of 4656 comment sentences extracted from closed issues of three OSS projects related to machine learning, labeled with the identified information types; and (3) evaluation of automated, supervised text classification mechanisms to identify the information types of sentences in OSS issue threads.


%% file: related_work.tex
\section{Related Work}
Our work is closely related to previous studies that focused on (1) the role of ITSs in software engineering activities and (2) classification of software-related texts. We briefly review the recent work in each category.

\subsection{Issue Tracking Systems}
The literature has established that ITSs contain rich information about the software project and play an important role in various software engineering activities. Heck and Zaidman \cite{Heck2017} examined the just-in-time requirements represented in ITSs and analyzed their quality against a framework they developed. Huang et al. \cite{Huang2017} proposed a machine learning approach in identifying software packages that may be affected by a certain issue based on the information contained in the issue report. Xia et al. \cite{Xia2017} utilized a topic modeling approach that leverages the issue descriptions to support bug triaging. Merten et al. \cite{Merten2016} utilized natural language processing (NLP) techniques to detect software feature requests in ITSs. While their technique was able to detect the feature request at the issue level with a satisfying quality, it was not effective at finding the exact sentence that included the issue request. Rastkar et al. \cite{Rastkar2014} have also investigated summarization techniques of issue reports that can help software developers quickly grasp the information recorded in ITSs. Based on tasks and needs, however, different types of OSS stakeholders may not be equally satisfied by the same summary. In our work, we thus focused on investigating categories of the rich information in the discussion threads to support the diverse needs of stakeholders. Our work can serve as the first step towards multifaceted summarization of issue discussions.

ITSs are also an important platform that supports communication and collaboration among the software developers and other stakeholders, facilitating them to form a coherent community. Bertram et al. \cite{Bertram2010} conducted a qualitative study of ITSs used by small, collocated software development organizations and found that, although the teams were collocated, the ITS still served as ``a focal point for communication and coordination for many stakeholders within and beyond the software team.'' They found that in the ITS, each issue was treated as a thread of chatroom conversations, with the issue report itself as the topic of the thread. Their findings also suggested that the ITS serves different needs to the software stakeholders and thus requires role-oriented data filtering mechanisms and interfaces. Comparing to collocated teams, OSS communities usually involve a larger number of participants that have a more diverse background, experience, expertise, and needs. Thus the role of the ITSs in supporting collaboration and community involvement is even more crucial in OSS communities.

\subsection{Classification of Software-Related Text}
Our work is most closely related to previous studies that focused on identifying information types or extracting topics from texts generated during software engineering activities. Similar to our work, Ko and Chilana \cite{Ko2011} have conducted a qualitative analysis of bug reports to identify the topics involved in the discussion threads. However, the purpose of their analysis is to understand the focus and the dynamics of such discussions, whereas our goal is to identify useful information types to satisfy the diverse needs of OSS stakeholders. Viviani et al. \cite{Viviani2018} have analyzed the design topics presented in the discussion of pull requests at the paragraph level in order to explore the types of design-related information. While we follow a similar manual annotation approach, our work focused on a sentence-level analysis of a wider range of information that goes beyond a design perspective.

Many researchers have also investigated automated techniques to classify software-related text. A majority of research in this area relies on a labeled corpus, upon which a supervised learning algorithm can be trained and evaluated. Consequently, manually identifying the topics that are comprised in the data under question is usually an integral first step in these studies.
For example, Panichella et al. \cite{Panichella2015} identified different topics presented in app store reviews in order to support developers to distinguish relevant and constructive feedback. They used a combination of text analysis and sentiment analysis techniques to categorize 1421 review sentences into four general topics. Alkadhi et al. \cite{Alkadhi2017} reported on an exploratory study aimed at identifying elements of rationales appeared in developer chat messages. They first conducted a content analysis to identify the frequency and completeness of the rationales and then explored automated techniques to classify the rationale elements. Similarly, Wood et al. \cite{Wood2018} differentiated 26 \textit{Speech Acts} (i.e., utterances that have an actional function) in chat conversations about bug repairs and developed an automated classifier to identify the speech acts. Also focused on speech acts, Morales-Ramirez et al. \cite{Morales-Ramirez2018, Morales-Ramirez2017} identified requirements-relevant information in OSS issue discussions using NLP and linguistic parsing techniques.



Our study is different from the previous work as we focused on identifying important information types presented in issue discussions that would be relevant for various kinds of OSS stakeholders and serve different purposes. These information types will help reduce the amount of time a user would take to navigate through discussion threads and retrieve the information relevant to his/her particular scenario. Additionally, we evaluate the effectiveness of using automated supervised classification techniques to identify the information types.

%% file: issue_comment_coding.tex
\section{Issue Comment Analysis}
\label{sec:comment_analysis}
In this section, we aim to answer \textbf{RQ1}: \textit{What are the main information types presented in issue discussions that can potentially satisfy the needs  of different stakeholders?} Particularly, we discuss our empirical study on information types in the issue discussions. This step allows us to (1) inductively identify the primary information types that OSS participants have focused on during issue discussions and (2) generate a dataset that includes 4656 comment sentences, labeled with the information types, to train and evaluate automatic models. During our study, we use the term \textbf{\textit{issue thread}} to indicate the complete discussion on the posted issue, \textbf{\textit{issue comment}} to indicate a single post written by one participant of the issue (for our purpose, the original post of the issue was also considered as an issue comment), and \textbf{\textit{sentence}} to indicate one sentence within a comment -- each comment may contain one or more sentences. Our study is at the granularity level of comment sentences.

\begin{table*}[h]
\centering
\begin{center}
\caption{Statistics of studied projects and selected issues}
\label{tab:ai_project_stats}
\begin{tabular}{|c|p{2cm}|c|c|c|l|c|}
\hline
Project Name & \multicolumn{1}{c|}{Description} & Language & Stars\# & \begin{tabular}[c]{@{}c@{}}Closed\\ Issues\#\end{tabular} & \multicolumn{1}{c|}{Selected Issue Title} &  Comments\# \\ \hline
\multirow{5}{*}{Tensorflow} & \multirow{5}{*}{\begin{tabular}[c]{@{}l@{}}Graph-based\\ library for\\ numerical\\ computations\end{tabular}} & \multirow{5}{*}{C++} & \multirow{5}{*}{104k} & \multirow{5}{*}{10888} & Node.js (JavaScript) Wrapper API & 246 \\ \cline{6-7}
 &  &  &  &  & Redesigning TensorFlow's input pipelines & 135 \\ \cline{6-7}
 &  &  &  &  & Easy to use batch norm layer & 128 \\ \cline{6-7}
 &  &  &  &  & ImportError: cannot open shared object file ... & 108 \\ \cline{6-7}
 &  &  &  &  & ValueError: Attempt to reuse RNNCell ... & 103 \\ \hline
\multirow{5}{*}{Scikit-learn} & \multirow{5}{*}{\begin{tabular}[c]{@{}l@{}}ML library for\\ data mining\\ and analysis\end{tabular}} & \multirow{5}{*}{Python} & \multirow{5}{*}{29.1k} & \multirow{5}{*}{4111} & t-SNE fails with array must not contain infs or NaNs & 107 \\ \cline{6-7}
 &  &  &  &  & GridSearchCV parallel execution with own scorer freezes & 98 \\ \cline{6-7}
 &  &  &  &  & Debian test failures ... & 85 \\ \cline{6-7}
 &  &  &  &  & Fitting additional estimators for ensemble methods & 74 \\ \cline{6-7}
 &  &  &  &  & Rethinking the CategoricalEncoder API & 64 \\ \hline
\multirow{5}{*}{SpaCy} & \multirow{5}{*}{\begin{tabular}[c]{@{}l@{}}Natural Language\\ Processing library\end{tabular}} & \multirow{5}{*}{\begin{tabular}[c]{@{}l@{}}Python/\\Cython\end{tabular}} & \multirow{5}{*}{9.8k} & \multirow{5}{*}{1683} & Use in Apache Spark / English() object cannot be pickled & 54 \\ \cline{6-7}
 &  &  &  &  & Streaming Data Memory Growth & 38 \\ \cline{6-7}
 &  &  &  &  & Additional Language Support & 37 \\ \cline{6-7}
 &  &  &  &  & pipe(): ValueError Error parsing doc & 25 \\ \cline{6-7}
 &  &  &  &  & Feature Request: Vector "File" interface & 24 \\ \hline
\end{tabular}
\end{center}
\vspace{-5pt}
\begin{tablenotes}
      \item \hspace{10pt}\textit{Note: all issue content are downloaded on July 4th, 2018}
    \end{tablenotes}
\vspace{-8pt}
\end{table*}

\subsection{Methodology}

\subsubsection{Data preparation}
To understand the multifaceted nature of the issue discussions for OSS projects, we focused our study on three Artificial Intelligence (AI) libraries hosted on GitHub, namely, TensorFlow\footnote{\url{https://github.com/tensorflow/tensorflow}}, scikit-learn\footnote{\url{https://github.com/scikit-learn/scikit-learn}}, spaCy\footnote{\url{https://github.com/explosion/spaCy}}. We chose those libraries for the following reasons. First, they are under active development. Since the beginning of each project to date, an average of 265, 62, and 40 commits are made per week for each project respectively. Secondly, large communities are formed around those projects during their development process. For example, there are more than 100 recognized contributors for spaCy, who made wide-ranging contributions including developing features, reporting issues, fixing bugs, conducting code reviews, and updating project documentation and websites. Finally, we chose those projects because the authors of this study have sufficient expertise on this domain to conduct a comprehensive and accurate analysis of the issue discussions in those projects. We selected the five most-commented, closed issues from each of the projects for analysis. We only included closed issues in the analysis because they allow us to understand the flow of a full discussion and can help to reveal all possible information types. However, the identified information types should also be applicable to open issues. The detailed characteristics of those projects and the selected issues are described in Table \ref{tab:ai_project_stats}.



\subsubsection{Qualitative Content Analysis Process}
We followed a qualitative content analysis process \cite{Merriam2015} to identify the information types for each sentence in the issue title, the original post, and all the other comments in the selected issue discussion threads. This process involved four steps:
\begin{enumerate}[label=\alph*.]
    \item Following an inductive approach, four annotators (i.e. authors of this paper) first independently coded selected discussion threads at the sentence level to identify the possible categories of comment sentences.
    \item The four annotators then met and exercised an affinity diagramming activity to group their individual codes. A code map was drawn linking different related codes from different annotators; duplicate codes were merged, and similar codes were grouped under a single high-level code. The meeting resulted in a preliminary codebook that contained the high-level codes and their descriptions.
    \item Using the initial codebook, two annotators then separately conducted a second round of coding of three randomly selected issues, one from each project. Their coding results were compared and their agreement was calculated using Cohen's kappa coefficient \cite{cohen1960coefficient}. The overall Kappa for the 1090 sentences coded in this step is 0.71, representing a substantial agreement between the two coders \cite{landis1977kappa_measurement}.
    They then discussed to resolve any disagreement and further refined the codebook.
    \item Finally, these two annotators separately re-coded the rest of the discussion threads using the final codebook.
\end{enumerate}

\subsection{Results}
\label{subsec:code_result}
Using a content analysis process, we identified a total of 54 information types in the issue discussion threads. We further grouped them into 16 major types that we describe here.

\textit{Expected Behaviour}, in which stakeholders discuss, from the user's perspective, the expected or ideal situation affected by the issue. This discussion sometimes relies on the personal preferences and opinions from the OSS participants. For example, a participant commented: ``My suggestion/request in the near term would be to have an option to make the vocabulary read only so that users who want to be able to leave spacy alone to do streaming data processing don't need to worry about changing memory requirements.''

\textit{Motivation}, in which stakeholders elaborate on why the issue needs to be fixed or a feature needs to be added. To strengthen their arguments, they usually described use cases involving the requested feature and/or cited competitors who implemented the requested feature. For example, in support of redesigning the TensorFlow's input pipeline one participant wrote: ``Right now, this method starves my GPU all the time, which is a shame because most other [deep learning] frameworks manage to make this much more performantly.''

\textit{Observed Bug Behaviour}, which only appears in bug reports and focuses on describing the observed behaviour of the bug. For example, one participant commented: ``I found strange behavior using the `pipe()' method'', then started to describe this behavior.

\textit{Bug Reproduction}, which also only appears in bug reports and focuses on any report, request, and/or question regarding the reproduction of the bug. For example, one participant commented that a bug was reproducible: ``Same problem here, working on Windows 10 with German text.''

\textit{Investigation and Exploration}, in which OSS stakeholders discuss their exploration of ideas about the problem that was thought to have caused the issue. Sometimes participants provide suggestions on how or what to investigate. For example, ``This result confirms my hypothesis but also shows that the memory increase really isn't all that significant... But it still points to a potential flaw in the design of the library.''

\textit{Solution Discussion} is framed around the solution space from the developers' point of view, in which participants discuss design ideas and implementation details, as well as suggestions, constraints, challenges, and useful references around such topics. For example, ``I know there are multiple ways of approaching this however I strongly recommend node-gyp for performance.''

\textit{Contribution and Commitment}, in which participants call for contributors and/or voice willingness or unwillingness to contribute to resolving the issue. For example, one potential collaborator said: ``I will gladly contribute in any way I can, however, this is something I will not be able to do alone. Would be best if a few other people is interested as well...''

\textit{Task Progress}, in which stakeholders request or report progress of tasks and sub-tasks towards the solution of the issue. Participants sometimes also mention their plan of actions. For example, ``I made an initial stab at it... - this is just a proof of concept that gets the version string into nodejs. I'll start working on adding the swig interfaces...''.

\textit{Testing}, in which participants discuss the testing procedure and results, as well as the system environment, code, data, and feedback involved in testing. For example, ``Tested on `0.101' and `master' - the issue seems to be fixed on `master' not just for the example document, but for the entire corpus...''

\textit{Future Plan}, in which participants discuss the long-term plan related to the issue; such plans usually involve work/ideas that are not required to close the current issue. For example, ``For the futures, stay tuned, as we're prototyping something in this direction.''

\textit{Potential New Issues and Requests}, in which participants identify and discuss new bugs or needed features while investigating and addressing the current issue. They are out of the scope of the discussion of the current issue but may lead to new issue reports. For example, when discussing a bug in scikit-learn about parallel execution that causes process hanging, one participant said: ``As a side point, I note there seems to be a lot more joblib parallelisation overhead in master ... that wasn't there in 0.14.''

\textit{Solution Usage} was usually discussed once a full or partial solution of the issue was released and stakeholders asked questions or provided suggestions about how to use the library with the new solution update. For example, ``Please help me how to continue training the model [with the new release].''

\textit{Workarounds} focus on discussions about temporary or alternative solutions that can help overcome the issue until the official fix or enhancement is released. In a discussion regarding memory growth for streamed data, one participant expressed his temporary solution: ``For now workaround with reloading / collecting nlp object works quite ok in production.''

\textit{Issue Content Management} focuses on redirecting the discussions and controlling the quality of the comments with respect to the issue. For example, ``We might want to move this discussion to here: [link to another issue]''.

\textit{Action on Issue}, in which participants comment on the proper actions to perform on the issue itself. For example, ``I'm going to close this issue because it's old and most of the information here is now out of date.''

\textit{Social Conversation}, in which participants express emotions such as appreciation, disappointment, annoyance, regret, etc. or engage in small talk. For example, ``I'm so glad that this has received so much thought and attention!''
\vspace{10pt}

\begin{figure}[!tbp]
\centering
\includegraphics[width=\columnwidth]{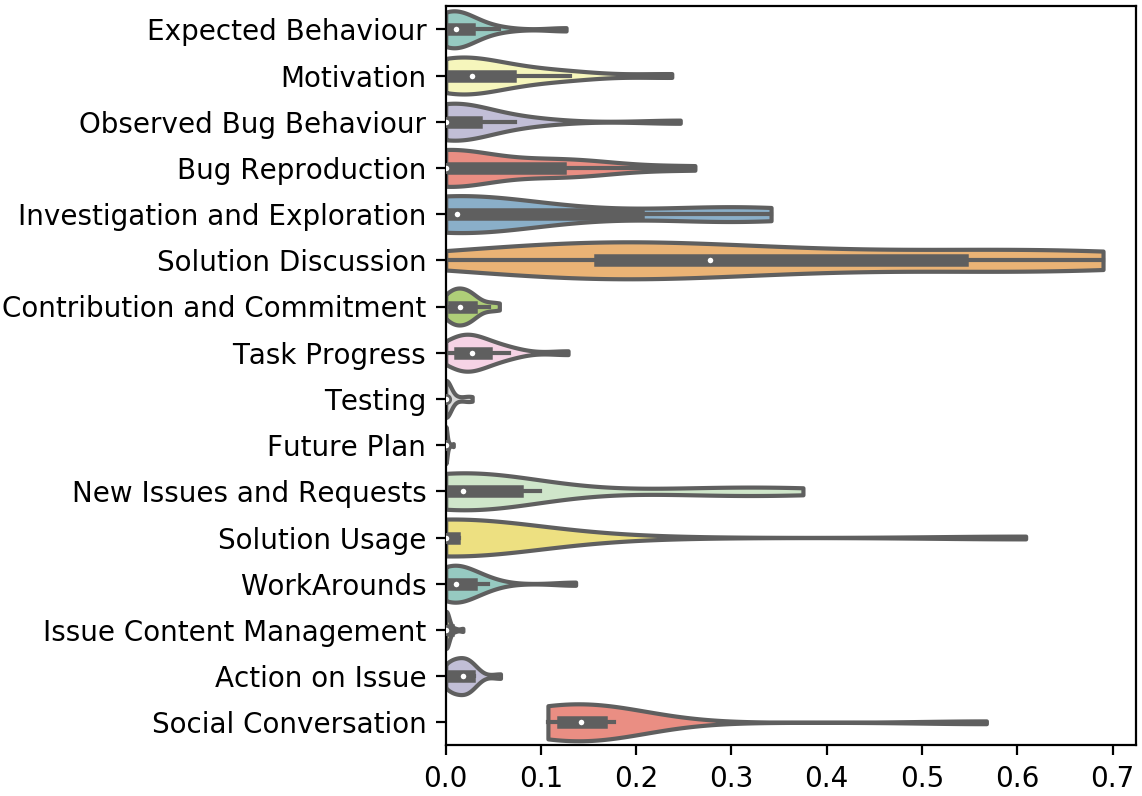}
\caption{Distribution of the percentage of the identified information types in the selected issues.}
\label{fig:code-distribution}
\end{figure}

The distribution of the percentage of the identified codes in the selected issues can be seen in Figure \ref{fig:code-distribution}. To assist future research in this area, we release our codebook that contains all the code themes, descriptions, and coding examples, as well as the coded dataset\footnote{\label{release_link}\url{https://git.io/fhQTt}}.
Because we focused on the longest and thus richest discussion threads of a diverse nature (i.e. bugs, feature requests, and configuration issues), we expect these codes would cover the most prominent types of information within issue discussion threads.
Among those types, some have already attracted research interests in the past, e.g. \textit{Observed Bug Behavior} and \textit{Solution Discussion}, for dealing with problems such as bug triage and impact analysis. However, many other types, such as \textit{Workarounds}, \textit{Future Plan}, \textit{Motivation}, and \textit{Solution Usage} are less studied but can be highly informative for stakeholders in various scenarios.

%% file: use_case.tex
\section{Use Case}
\begin{figure*}[tbp]
    \centering
    \begin{subfigure}[b]{0.45\textwidth}
        \centering
        \includegraphics[width=\linewidth]{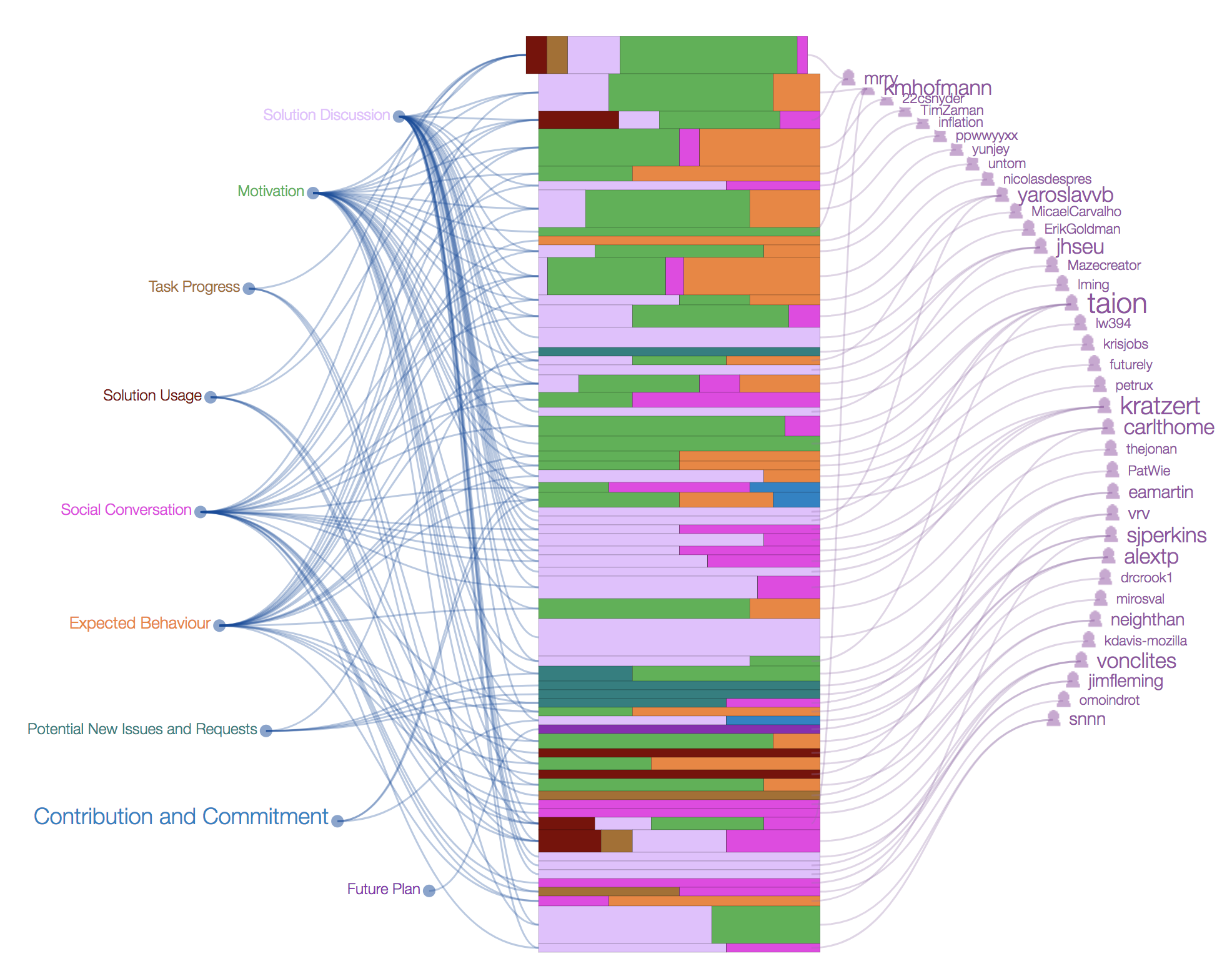}
        \caption{Full visualization}
        \label{fig:viz_a}
    \end{subfigure}%
    ~
    \begin{subfigure}[b]{0.45\textwidth}
        \centering
        \includegraphics[width=\linewidth]{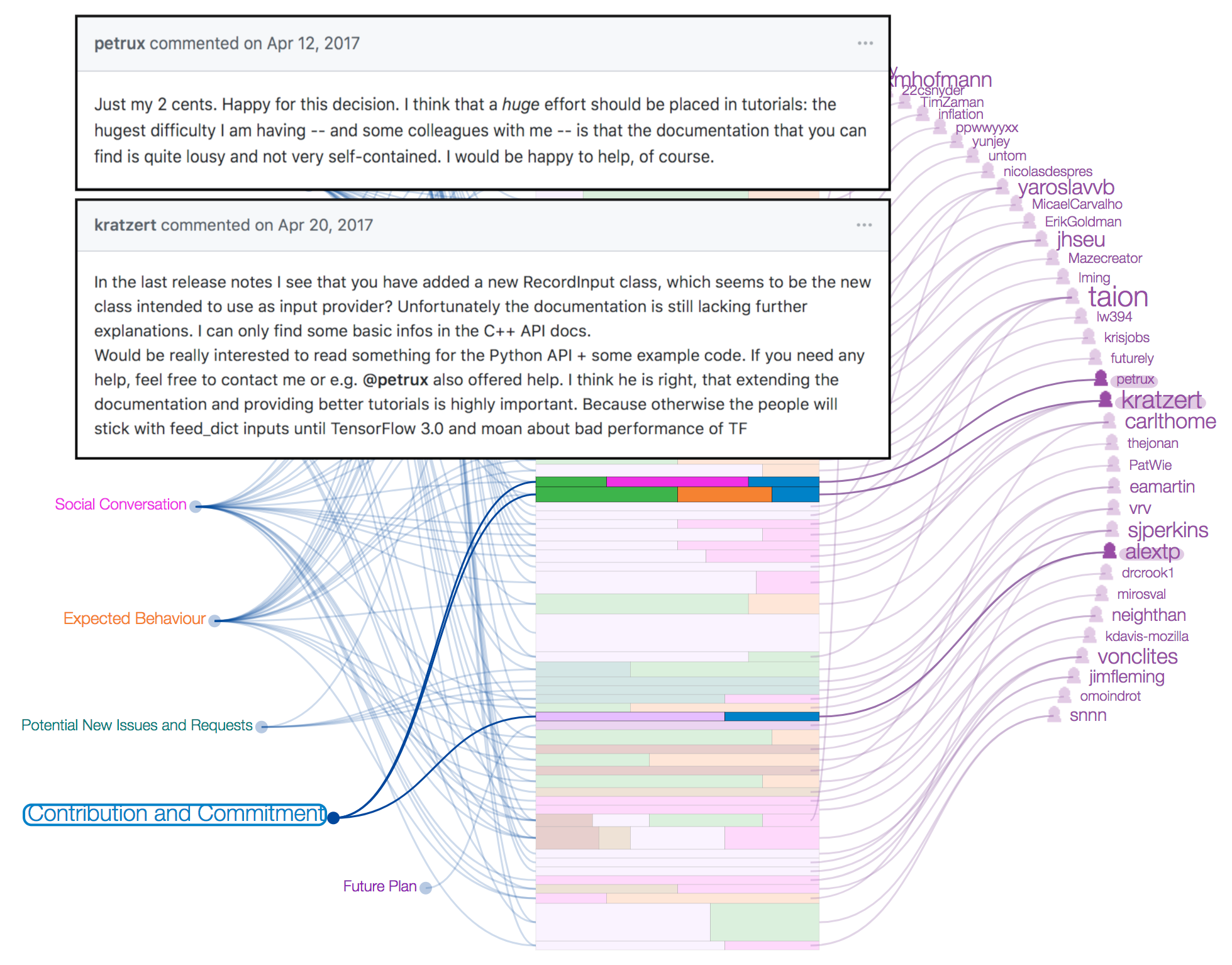}
        \caption{Focused on \textit{Contribution and Commitment}}
        \label{fig:viz_b}
    \end{subfigure}
    \caption{Visualization of a discussion thread \textit{TensorFlow \#7951} from beginning to May 20, 2017 with information types}
\end{figure*}

In this section, we present a use case demonstrating how OSS participants would be able to extract useful information from the discussion threads with the support of the information types we identified. To illustrate the use cases, we used the ConVis visualization tool \cite{convis2014}, which was originally developed to support interactive exploration of blog comments. This tool provides visual cues that separate parts of each comment into high-level topics/types and is thus suitable for our purpose. Figure \ref{fig:viz_a} shows a full visualization of an issue thread. In the center of the visualization, each horizontal bar represents a comment, with its height proportional to the comment length. The information types are color-coded and appear both on the comment bars (area representing length) and on the left side of the visualization. The right side indicates the participants who made the comments.

We use TensorFlow \textit{Issue \#7951 Redesigning TensorFlow's input pipelines} as an example. This issue was reported on February 28, 2017, and closed on August 30, 2017, accumulating 134 comments. We suppose a user accessed the thread on May 20, 2017, when the issue was still open and 65 comments were made on it, and consider the following use case:

A TensorFlow user who was resolving some difficulty importing and using data in her TensorFlow program had found this thread. Using the visualization tool supported by the information types, she quickly identified several comments that provided a lot of information on the \textit{Motivation} of redesigning the input pipeline and some \textit{Expected Behaviours}, by core project contributors and other users. She was glad that the TensorFlow team was actively working on this issue. However, she found that there were still participants who were willing to contribute (\textit{Contribution and Commitment}, Figure \ref{fig:viz_b}); after looking at those comments, she realized that these participants have proposed contributions to tutorials and documentation. She agreed that the current tutorial on importing data needs improvements. Based on her recent experience dealing with the issues, she had some insights to ameliorate it. So she decided to leave a comment in this thread to express her contribution willingness too. Looking at the \textit{Potential New Issues and Requests}, she found that one contributor had proposed to allow numpy to share buffers with TensorFlow variables. She found this proposition would be helpful in addressing her difficulty too and decided to follow up. While disappointed to see that there are currently no \textit{Workarounds} available, she decided to look at the \textit{Task Progress}. To her surprise, a feature that allowed a user to switch among train, validation, and test datasets at run time had been developed, which partially resolved her challenge.

Hence, with the help of the information types and some navigation support, OSS participants could retrieve and discover information that is otherwise buried in the discussion threads.

%% file: supervised_classification.tex
\section{Automated Information Type Detection}
Manually labeling the comments at sentence level is a time-consuming task, especially for long discussions. Therefore, automated detection of the information types would be critical for designing ITSs tools leveraging these types, such as information navigation, retrieval, and summarization tools. Our initial exploration suggests that text-based unsupervised methods such as topic modeling and clustering are not effective in detecting the information types. The performances of such methods are sensitive to the parameter settings \cite{agrawal2018wrong} and are highly dependant on the distribution of the terms which can not be generalized across domains \cite{chen2016survey}. Consequently, in this section, we explore the possibility of utilizing supervised techniques to detect information types of sentences in issue comments and the effectiveness of such techniques on each information types.

\begin{figure}[!tb]
    \centering
    \includegraphics[width=0.7\columnwidth]{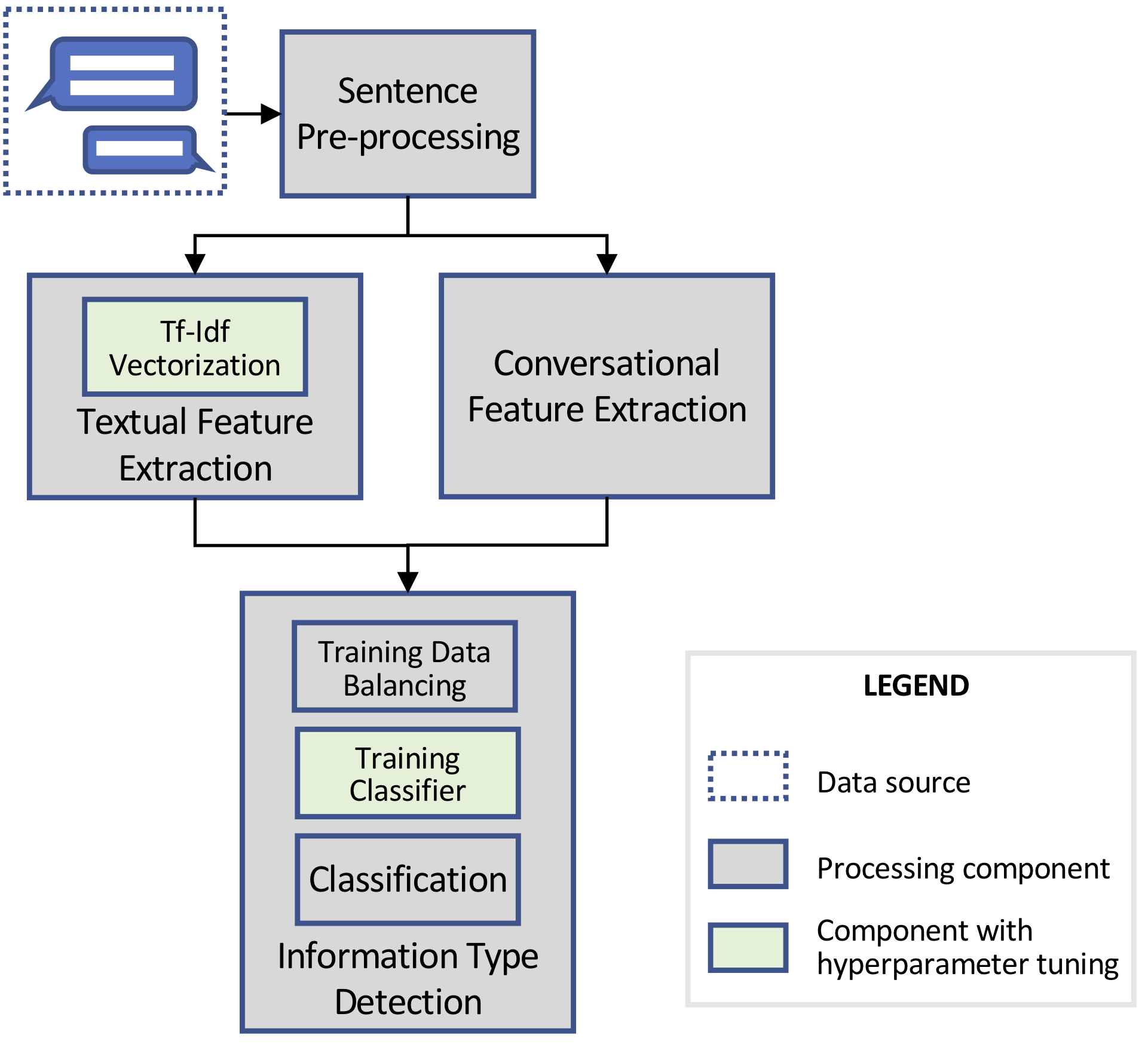}
    \caption{The process of supervised information type detection}
    \label{fig:supervised_process}
    \vspace{-8pt}
\end{figure}

\addtolength{\tabcolsep}{-2pt}
\begin{table*}[!t]
\centering
\caption{List of conversational features}
\label{tab:features}
\begin{tabular}{|c|c|l|c|}
\hline
Feature Type & Feature Name & \multicolumn{1}{c|}{Description} & Value Range \\ \hline
\multirow{2}{*}{Participant} & AA & Author's association with repository. & \{OWNER, CL, MBR, OTHER\}\\
& BEGAUTH & Flag of whether the comment author also posted the original issue. &\{True, False\}\\ \hline
\multirow{3}{*}{Length} & LEN & Length of the sentence in terms of character count.& \{Positive Numbers\} \\
& TLEN & Count of words in sentence divided by that of the longest sentence in thread & (0, 1] \\
& CLEN & Count of words in sentence divided by that of the longest sentence in comment. &(0, 1] \\ \hline
\multirow{4}{*}{Structural} & TLOC & Position of sentence in comment divided by the number of sentences in comment.& (0, 1]\\
& CLOC & Position of sentence in conversation divided by the number of sentences in thread &(0, 1]\\
& FIRST\_TURN & Flag of whether if this is in the first comment. &\{True, False\}\\
& LAST\_TURN & Flag of whether this is the last comment or not & \{True, False\}\\ \hline
\multirow{4}{*}{Temporal} & TPOS1 & Time from beginning of conversation to comment divided by the total time of thread. &[0, 1]\\
& TPOS2 & Time from comment to end of conversation divided by the  total time of thread. &[0, 1]\\
& PPAU & Time from previous comment to current comment (normalized). & [0, 1]\\
& NPAU & Time from current comment to next comment (normalized). & [0, 1]\\ \hline
Code & HAS\_CODE & Flag to indicate whether the comment contains a code snippet.& \{True, False\} \\\hline
\end{tabular}
\begin{tablenotes}
      \item \textit{Note: CL - Collaborator, MBR - Member}
\end{tablenotes}
\end{table*}

Supervised information type classification techniques leverage the labeled dataset we created in Section \ref{sec:comment_analysis}. The process of the supervised classification is comprised of three main steps: sentence pre-processing, feature extraction, and information type detection, as shown in Figure \ref{fig:supervised_process}. We describe the key components and their potential configurations during each step in the rest of this section.

\subsection{Sentence Pre-processing}\label{subsec:data_preprocessing}

Issue comments are generally noisy for automated textual classification tasks. In order to clean the sentences in the issue comments, we took the following steps:
\begin{enumerate}
    \item Identify embedded source code blocks and replace with \textbf{CODE} token;
    \item Identify quotations made in previous comments and replace with \textbf{QUOTE} token;
    \item Identify reference links to external resources and replace with \textbf{URL} token; and
    \item Identify mentions to GitHub users and replace with \textbf{SCREEN\_NAME} token.
\end{enumerate}
All the above identification and replacement were achieved by using regular expressions to match the formatting syntax of GitHub issues.

We then tokenized each sentence into words, lemmatized and lowercased each word, and then removed punctuation and common contractions. However, we preserved the stop words because a common stop-word list might filter out important terms in domain-specific and task-specific context. Our exploratory investigation revealed that certain words could contribute to sentence semantics relevant to the classification of information types. For example, the word ‘please’ can be a strong indicator for the information type of \textit{Action on Issue}.

\subsection{Feature Extraction}\label{subsec:feature_extraction}
We considered two sets of features to characterize the sentence in issue comments: \textit{Textual Features} and \textit{Conversational Features}. Here, we introduce the two types of feature sets:

$\bullet$ \textit{Textual Features} are extracted from the textual content of each individual sentence. Each word in the sentence after the pre-processing process acts as one feature. We also use n-grams as additional features which represent the appearance of $n$ tokens sequences. The features are then transformed from the sentence into numerical representation using the TF-IDF weighting method in which the frequencies of words and n-grams in the sentence are multiplied by their inverse document-frequency.

$\bullet$ \textit{Conversational Features} focus on the characteristics that describe the conversational context in which each sentence situates during the issue discussion. They can be further divided into the following groups: \textbf{Participant Features} describe the role that the sentence author has in the project (i.e. owner, collaborator, member, or other) and in the current issue thread (i.e. original issue author or not); \textbf{Length Features} depict absolute length of the sentence and relative length with respect to other sentences in the issue comment and in the thread; \textbf{Structural Features} describe the location of the sentence in relation to the whole discussion thread; \textbf{Temporal Features} describe the time when the comment is made in relation to the immediately previous and next comment, and in relation to the whole discussion thread; and finally \textbf{Code Feature} indicates if the current issue comment contains code snippet. The above conversational features were inspired by previous work on summarizing bug report \cite{rastkar2010summarizing}, and are summarized in Table~\ref{tab:features}.

\subsection{Information Type Detection}\label{subsec:type_detection}

\subsubsection{Balancing Training Data}
After extracting features from the sentence, we used supervised classification techniques to detect information types. As depicted in Figure \ref{fig:code-distribution}, the distribution of different information types is highly imbalanced. In order to effectively train the classifier in a manner such that it is not biased by the sentences from the majority types, we first needed to perform balancing techniques to the data that would be used for training the classifier. We explored the following two techniques:

$\bullet$ \textit{Adjusting class weight} can be performed during the training process, i.e. increasing the importance of sentences from minority types. This technique penalizes mistakes in samples differently based on the number of samples in each class.

$\bullet$ \textit{SMOTE}, or Synthetic Minority Over-Sampling Technique, is a method that combines over-sampling the minority class by creating new samples and under-sampling the majority class in order to produce a training dataset that improves classifier performance \cite{Chawla2002}. It has been proven effective in handling imbalance datasets in many software engineering tasks \cite{Kamei:2007:EOU:1302496.1302937, pelayo2007applying} and classification tasks elsewhere \cite{Chawla2002}.

\vspace{5pt}

\subsubsection{Training Classifier}
Two classifiers are considered for detecting the information type of sentences, i.e. \textit{Logistic Regression} and \textit{Random Forest}. These two classifiers are commonly used in text classification tasks \cite{forman2003extensive, adler2011wikipedia}. We discarded other classifiers in our experiments, such as Support Vector Machine and Naive Bayes, because they yielded inferior results and took a substantially longer time to train during our initial investigation.

$\bullet$ \textit{Logistic Regression} estimates the probability of an event or class based on a linear combination of the input features. It is essentially a binary classifier, but it can be generalized to a multi-class logistic regression classifier using a one-vs-rest scheme in which the model treats each label as a binary classification problem (i.e. the samples with the target label being one class and the samples with all the other labels as being another class).

$\bullet$ \textit{Random Forest} is an ensemble machine learning algorithm that constructs multiple decision trees based on the provided training data. The categorized class is then the most frequent label identified by the trees from the forest.

%% file: evaluation.tex
\section{Evaluation of Automated Detection Methods}

In this section, we aim to address \textbf{RQ2}: \textit{To what extent can automated classification methods identify the information types of issue comments?} and \textbf{RQ2.1} \textit{What are the performance and trade-offs of using different methods}?  We first describe how we designed the experiments to evaluate the performance of different configurations of the proposed automated information type detection methods for varied scenarios. We then discuss our observations on the evaluation results and the implications when applying those methods in practice.

\subsection{Experiment Design}

We utilized the sentences that were annotated during the coding process (see Section \ref{subsec:code_result}) to evaluate the information type detection techniques. Out of the 4656 annotated sentences obtained, 293 have multiple labels. In addition, three labels appear only in 33 sentences altogether, representing less than 1\% of the total dataset. These labels are \textit{Future Plan},  \textit{Content Management} and  \textit{Testing-Related}. For our experiments, we ignored those sentences because the sample size would be insufficient to train the classifiers effectively. We performed the experiment using the remaining 4330 annotated sentences.

We considered two evaluating scenarios where the automated methods can be used in realistic settings:\\
$\bullet$  \textit{Scenario 1}: The sentences in the discussion are partially categorized with information types and the users want to retrieve the missing labels in the same discussion. \\
$\bullet$ \textit{Scenario 2}: Comments in a new discussion thread need to be categorized, given knowledge about other threads.

Accordingly, we designed two series of experiments to evaluate the performance of the classifiers for the above scenarios: (1) Stratified 5-fold cross validation and (2) Leave-One-Issue-Out cross validation. In particular, in the first series of experiments, the sentence samples were partitioned into five equal-sized subsamples with stratification (i.e. splitting the data such that roughly the same percentage of information type labels lie in each partition); stratification was used to ensure that the original distribution of information types was preserved in each partition. Every time, four folds of data were used to train the classification models and the remaining one fold was used for testing. This process iterated 5 times until every fold was used for testing once. In the second series of experiment, we used all the sentences from 14 issue threads to train the classifier, and all the sentences from the left-out issue for testing. This process iterated 15 times in total until every issue was tested once. The second series of experiments do not guarantee that the distributions of the training and testing data are similar, but they are closer approximations of Scenario 2.

In each series of experiments, three independent variables were included to represent all the possible configurations of the detection methods. First, the two classification models, i.e., Logistic Regression (LR) and Random Forest (RF) were examined. Second, we incorporated three types of feature sets: Textual Features only, Conversational Features only, and Textual + Conversational Features. Finally, the two imbalanced data handling techniques were included. Thus in total, each series of experiments contained 12 configurations. With these experiments, we intended to find the best model/technique configurations for the two evaluation scenarios. All possible configurations are summarized in Table \ref{tab:experimental_conditions}.

\begin{table}[!t]
\centering
\caption{All configurations for detecting information types}
\label{tab:experimental_conditions}
\begin{tabular}{clll}
\hline
\multicolumn{1}{c}{ID} & \multicolumn{1}{c}{Model} & \multicolumn{1}{c}{Feature Set} & \multicolumn{1}{c}{\begin{tabular}[c]{@{}c@{}}Imbalance Handling\end{tabular}} \\ \hline
LTC & Logistic Regression & Textual & Class Weight \\
LTS & Logistic Regression & Textual & SMOTE \\
LCC & Logistic Regression & Conversational & Class Weight \\
LCS & Logistic Regression & Conversational & SMOTE \\
LBC & Logistic Regression & Both & Class Weight \\
LBS & Logistic Regression & Both & SMOTE \\
RTC & Random Forest & Textual & Class Weight \\
RTS & Random Forest & Textual & SMOTE \\
RCC & Random Forest & Conversational & Class Weight \\
RCS & Random Forest & Conversational & SMOTE \\
RBC & Random Forest & Both & Class Weight \\
RBS & Random Forest & Both & SMOTE \\
\hline
\end{tabular}
\end{table}

\addtolength{\tabcolsep}{-2.5pt}
\begin{table*}[htb]
\centering
\caption{Searching space for hyperparameter tuning}
\label{tab:hyperparameters}
\begin{tabular}{|c|c|l|c|}
\hline
Component & Hyperparamater &  \multicolumn{1}{c|}{Description} & Values Searched \\ \hline
TF-IDF Vectorization  & ngram\_range & The lower and upper values of ngrams to be considered during the vectorization process & \{$(1,1)$, $(1,2)$\} \\ \hline
Logistic Regression & C & Inverse of the regularization strength where smaller values mean stronger regularization & \{$0.01, 0.1, 1, 10$\} \\ \hline
\multirow{2}{*}{Random Forest} & min\_samples\_split & Minimum number of datapoints required to split a node & \{2, 5, 10\}\\
& n\_estimators & Number of trees in the forest & \{10, 50, 100\} \\ \hline
\end{tabular}
\end{table*}

For certain components during the information type detection, parameters of the components need to be set before the learning process begins. Such parameters are called \textbf{hyperparameters} and the process of determining a good set of hyperparameters is called \textbf{hyperparameter tuning}. Hyperparameter tuning is essential for the machine learning model to be optimally trained for a specific problem. It can potentially lead to largely improved results \cite{fu2016tuning}. Therefore, we performed hyperparameter tuning in each condition for each scenario. The strategy we adopted is called nested cross-validation. After we split the dataset for training and testing, we again performed a stratified 5-fold validation (the inner step of nested cross validation). This technique further splits the training dataset into training (for fitting the model) and validation (for selecting the hyperparameters). The best hyperparameters were then chosen based on their average performance across the five cross validations and used to train and test the model. The components requiring hyperparameter tuning are highlighted in Figure \ref{fig:supervised_process}. The ranges of values for each hyperparameter are summarized in Table \ref{tab:hyperparameters}.

To evaluate our classifier, we used the following metrics: (1) \textit{Precision}, which refers to the ratio of the number of correct categorizations of a class to the total number of categorizations made of that class; (2) \textit{Recall}, which refers to the ratio of the number of correct categorizations of a class to the total number of data points in that class in the golden test set; and (3) \textit{F1-Score}, which is the harmonic mean of precision and recall. For each fold, we calculated the weighted average of the metrics of all classes (i.e. information types) to represent the overall quality of the model. The weighting is decided by \textit{Support}, which is the frequency of each class in the test set. We then used the weighted average of the F1-Scores of all folds to compare the different configurations for both evaluating scenarios. All our source code for the experiments are released along with the codebook and the annotated issue discussions to promote reproducibility and reuse (see Section \ref{subsec:code_result}).

\subsection{Results and Discussion}
\subsubsection{Scenario 1}

\begin{figure}[!b]
    \centering
    \includegraphics[width=0.99\columnwidth]{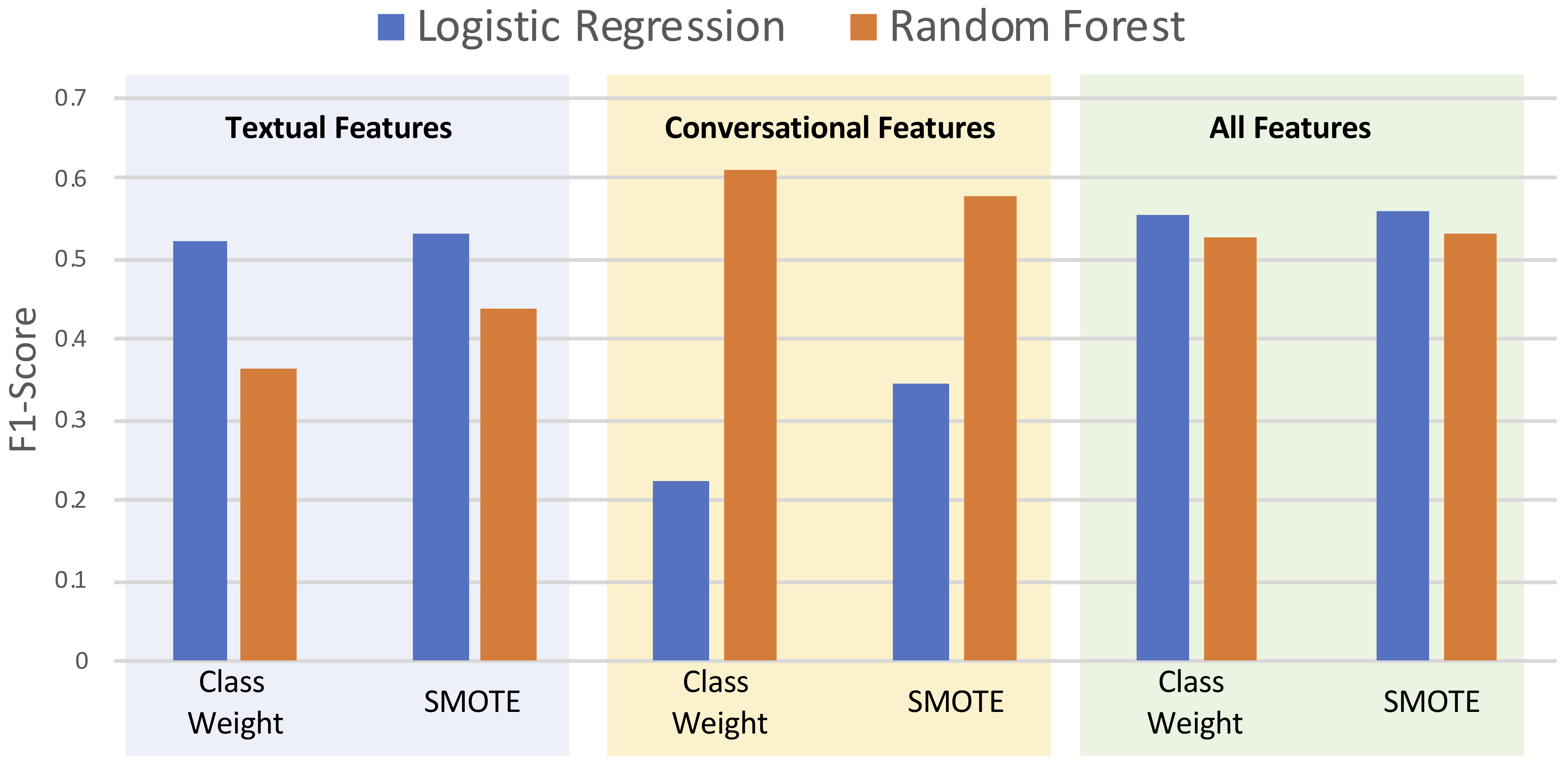}
    \caption{Comparison of F1-scores in Scenario 1 (Stratified  5-fold cross validation)  }
    \label{fig:5_fold}
\end{figure}

The F1-scores of the 12 configurations for Scenario 1 (i.e. Stratified 5-fold cross validation) are summarized in Figure \ref{fig:5_fold}. The overall best performance was achieved when using configuration RCC, i.e. Random Forest using Conversational features with class weight adjustment. The weighted average F1-score is 0.61. It is interesting to note that for Random Forest, using only conversational features yielded higher F1-score than using both textual and conversational features. One possible explanation is that Random Forest cannot effectively handle the sparsity of the textual features in our dataset. There are more than 5000 textual features generated by unique words from the dataset, and more than 30000 if adding bi-grams. Many of those features are uninformative about specific classes and result in weak trees being created during training.  Therefore the trained Random Forest classifiers did not generalize well to the testing data.

The Precision, Recall and F1-score of RCC on each information type are summarized in Table \ref{tab:5F_rcc_results}. We observed that the automated detection achieved satisfactory results on most of the information types, especially on \textit{Solution Usage} with a precision of 0.65 and a recall of 0.82. This observation reveals that conversational features, such as who made the comment, when the comment was made, and how long the sentence was, are strong indicators for most information types. On the other hand, the automated detection performed poorly on \textit{Expected Behaviour}, \textit{Contribution and Commitment}, and \textit{Task Progress}. These results imply that 
conversation on those types is more likely to appear throughout the issue thread with various lengths. The participants in the conversation are also likely to be from different role groups.

\begin{table}[!t]
\centering
\caption{Detailed results for each information type in Scenario 1 with configuration RCC (Random Forest using Conversational features with class weight adjustment)}
\label{tab:5F_rcc_results}
\begin{tabular}{ l c c c c }
 \hline
 \multicolumn{1}{c}{Label} & Precision & Recall & F1-Score & Support \\
 \hline
Expected Behaviour & 0.42 & 0.28 & 0.33 & 124 \\
Motivation & 0.56 & 0.53 & 0.54 & 288 \\
Observed Bug Behaviour & 0.56 & 0.70 & 0.62 & 131 \\
Bug Reproduction & 0.53 & 0.47 & 0.50 & 245 \\
Investigation and Exploration & 0.60 & 0.65 & 0.62 & 377 \\
Solution Discussion & 0.68 & 0.71 & 0.69 & 1411 \\
Contribution and Commitment & 0.25 & 0.19 & 0.21 & 83 \\
Task Progress & 0.27 & 0.14 & 0.18 & 125 \\
Potential New Issues and Requests & 0.67 & 0.66 & 0.66 & 230 \\
Solution Usage & 0.65 & 0.82 & 0.73 & 368 \\
Workarounds & 0.58 & 0.45 & 0.49 & 89 \\
Action on Issue & 0.45 & 0.39 & 0.42 & 61 \\
Social Conversation & 0.63 & 0.62 & 0.63 & 798 \\
 \hline
Weighted average/Total & 0.61 & 0.62 & 0.61 & 4330 \\
 \hline
\end{tabular}
\end{table}


\subsubsection{Scenario 2}
\begin{figure}[!t]
    \centering
    \includegraphics[width=0.99\columnwidth]{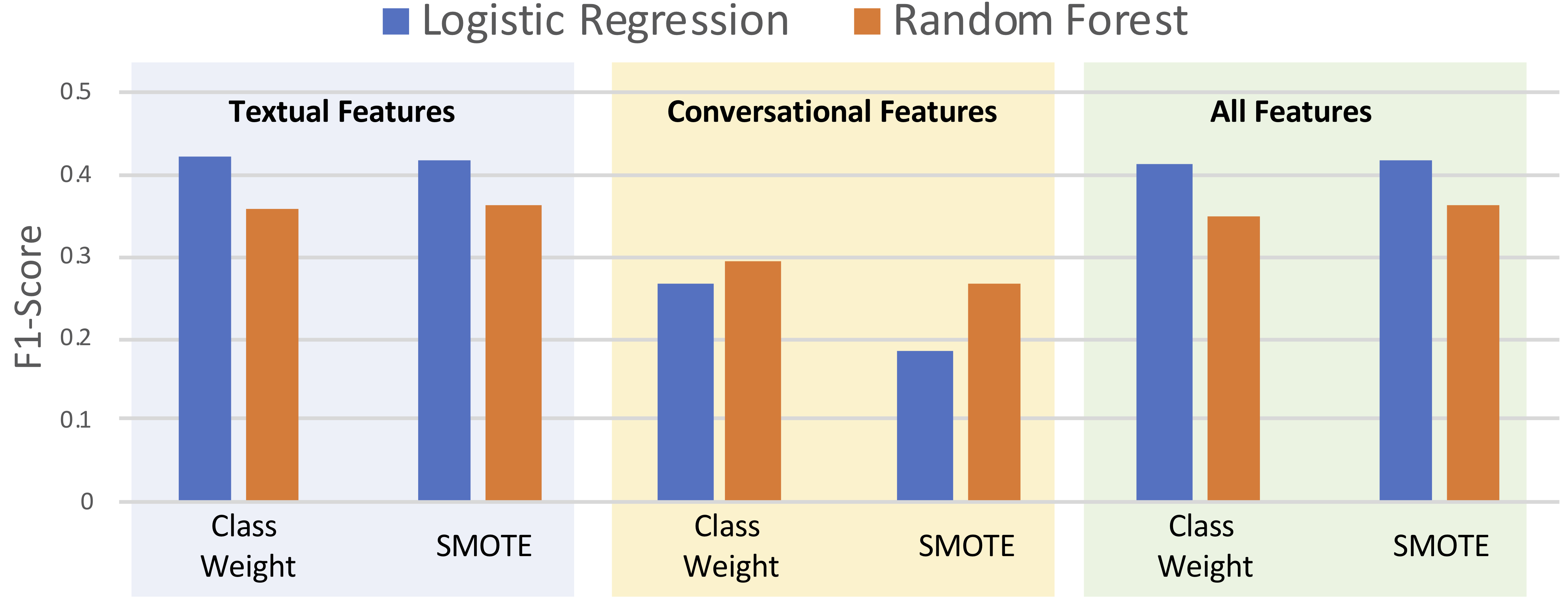}
    \caption{Comparison of F1-scores in Scenario 2 (Leave-One-Issue-Out cross validation)  }
    \label{fig:LOIO}
\end{figure}

As depicted in Figure \ref{fig:LOIO}, the results of Leave-One-Issue-Out cross validation are different from the 5-fold cross validation, indicating that when the relationship between training and testing data changes, the performance of information type detection also alters. Here, the best configuration is LTC, i.e. Logistic Regression using textual features with adjusting class weight, achieving the weighted average F1-score of 0.42. The per-label precision, recall and F1 scores are detailed in Table \ref{tab:loio_ltc_results}.

Different from Scenario 1, in Scenario 2 using only textual features and using both sets of features perform better than using conversational features for both classifiers. Such a finding implies that textual features are more reliable at detecting information types for sentences from new issues. Additionally, the average performance of all configurations in Scenario~2 is inferior to that in Scenario 1. One reason for these results may be the possible large variance of the conversation lengths and styles across different issues. Furthermore, results in Table \ref{tab:loio_ltc_results} indicate that sentences of some types, such as \textit{Social Conversation} and \textit{Action on Issue}, bear greater textual similarity across issues than others. In contrast, terms used in \textit{Observed Bug Behaviour}, \textit{Potential New Issues and Requests} and \textit{Work-around} are more likely to be context specific and can hardly be generalized to new issues.

\vspace{10pt}
In both scenarios, the F1-scores varied greatly depending on which classifier and feature set were used. For example, Random Forest outperforms Logistic Regression when using only conversational features. The techniques for handling imbalanced dataset, however, only yielded slightly different results. These findings clearly demonstrated the strengths of different classifiers for handling different types of features. Through this exploration, we emphasize the necessity of careful model selection and hyperparameter tuning to ensure the best performance when detecting information types.

\begin{table}[!t]
\centering
\caption{Detailed results for Scenario 2 with configuration LTC (Logistic Regression using textual features with class weight adjustment)}
\label{tab:loio_ltc_results}
\begin{tabular}{ l c c c c }
 \hline
 \multicolumn{1}{c}{Label} & Precision & Recall & F1-Score & Support \\ \hline
Expected Behaviour & 0.71 & 0.1 & 0.15 & 124 \\
Motivation & 0.44 & 0.1 & 0.13 & 288 \\
Observed Bug Behaviour & 0.23 & 0.03 & 0.04 & 131 \\
Bug Reproduction & 0.53 & 0.36 & 0.42 & 245 \\
Investigation and Exploration & 0.47 & 0.24 & 0.31 & 377 \\
Solution Discussion & 0.59 & 0.65 & 0.58 & 1411 \\
Contribution and Commitment & 0.51 & 0.31 & 0.37 & 83 \\
Task Progress & 0.35 & 0.26 & 0.29 & 125 \\
Potential New Issues and Requests & 0.1 & 0.03 & 0.03 & 230 \\
Solution Usage & 0.57 & 0.08 & 0.12 & 368 \\
Work-Arounds & 0.51 & 0.06 & 0.09 & 89 \\
Action on Issue & 0.78 & 0.49 & 0.58 & 61 \\
Social Conversation & 0.74 & 0.69 & 0.70 & 798 \\ \hline
Weighted average/Total & 0.55 & 0.42 & 0.42 & 4330 \\
\hline
\end{tabular}
\end{table}

%% file: threats_to_validity.tex
\section{Threats to Validity}
There are two primary threats to the validity of this study. First, due to the challenge and effort involved in manual annotation of the issue discussion threads, our work only analyzed 15 issues from three OSS projects. However, our intentional focus on the most-commented issues in the dynamic machine learning projects that involve diverse participants allowed us to analyze representative situations where support in obtaining information from issue discussion threads is most needed. Still, the characteristics of the selected projects and issues may have affected the information types we identified and the performance of the automated techniques, making our approach more or less effective. As a result, we cannot claim generalizability. In the future, we will apply our approach to a larger number of OSS projects and issues in order to evaluate the external validity of our results. We also plan to examine our approach on analyzing pull request discussions. 

Second, the interpretation of the issue comments and the initial categorization of the information types may subject to biases from the personal and professional experiences of the researchers. To mitigate this threat, we adopted a rigorous inductive analysis process that involved multiple rounds of coding and discussion. The overall Kappa score (0.71) of the second round of independent coding demonstrated the reliability of our codebook. Arguably, we are still not able to guarantee that the corpus we labeled with the information types is 100\% correct. However, we have made our best effort to create a high-quality dataset. In fact, data quality is a common issue in software engineering research, especially when examining human activities. As a future work, we plan to involve external researchers and OSS community participants to review our dataset in order to reinforce its quality.

%% file: conclusion.tex
\section{Conclusion}
In this paper, we have examined the information types contained in OSS issue discussions. Using a qualitative content analysis approach, we have identified 16 categories of information that can potentially support OSS participants retrieve and discover useful and otherwise hidden elements from the discussion threads. We further investigated the effectiveness of using supervised, automated techniques to classify the issue discussion sentences into the information types we identified. Our findings indicated that supervised classifiers such as Random Forest can effectively detect most sentence types using only conversational features when prior knowledge about the issue discussion is available. Logistic Regression methods can yield satisfactory performance using textual features when classifying sentences from new issues, particularly for certain information types such as \textit{Solution Discussion}, \textit{Action on Issue}, and \textit{Social Conversation} while falling short on others. 

Our work represents a nontrivial first step towards tools and techniques for identifying and obtaining the rich information recorded in ITSs. The wide-ranging information types identified in this work will help motivate software engineering research to leverage the various types of information from the issue discussion threads, a focus that is currently underrepresented in the literature. For example, the \textit{Solution Usage} and \textit{Workarounds} information types can be used to help generate or improve software documentation; the \textit{Motivation} information can potentially be used in competitive analysis and prediction of user drop-off. Additionally, the findings from this work can help realize unique issue thread navigation tools and context-sensitive issue summary techniques. In future work, we will investigate the design of such tools and conduct empirical studies to understand how the identified information types and tools can help diverse OSS participants more effectively utilize the ITSs. Additionally, we plan to explore the potential of our technique in aggregating information from a large number of issues posted and updated daily.